\documentclass[10pt,a4paper]{article}
\usepackage[utf8]{inputenc}
\usepackage{amssymb}
\usepackage[protrusion=true]{microtype}
\usepackage{verbatim} 
\usepackage{threeparttable} 
\usepackage{nicefrac}
\usepackage{graphicx} 
\usepackage[hidelinks]{hyperref} 
\usepackage{authblk}
\usepackage{booktabs}
\usepackage{caption}
\usepackage{subfig}
\usepackage{float}
\usepackage{multirow}
\usepackage{siunitx}

\usepackage[total={6in,8in},top=1in,bottom=1in]{geometry}

\usepackage[sectionbib,sort&compress,merge,numbers]{natbib}
\newcommand*{\indentintable}{\hspace*{0.2cm}}

\begin{document}

\title{$\alpha$-Indirect Control in Onion-like Networks}

\author[1]{Kirill Polovnikov\footnote{\texttt{\href{mailto:k.polovnikov@skoltech.ru}{k.polovnikov@skoltech.ru}}}}
\author[2]{Nikita Pospelov}
\author[3]{Dmitriy Skougarevskiy}

\affil[1]{\small Skolkovo Institute of Science and Technology, 121205 Moscow, Russia}
\affil[2]{\small Institute for Advanced Brain Studies, Lomonosov Moscow State University, 119234 Moscow, Russia}
\affil[3]{\small European University at St.~Petersburg, 191187 St.~Petersburg, Russia}


\maketitle

\begin{abstract}
    \noindent Tens of thousands of parent companies control millions of subsidiaries through long chains of intermediary entities in global corporate networks. Conversely, tens of millions of entities are directly held by sole owners.
    We~propose an~algorithm for identification of ultimate controlling entities in such networks that unifies direct and indirect control and allows for continuous interpolation between the two concepts via a factor damping long paths. By exploiting onion-like properties of ownership networks the algorithm scales linearly with the network size and handles circular ownership by design. We apply it to the universe of 4.2 mln UK companies and 4 mln of their holders to understand the distribution of control in the country. Furthermore, we provide the first independent evaluation of the control identification results. We reveal that the proposed $\alpha$-ICON algorithm identifies more than 96\% of beneficiary entities from the evaluation set and supersedes the existing approaches reported in the literature. We refer the superiority of $\alpha$-ICON algorithm to its ability to correctly identify the parents long away from their subsidiaries in the network.
\end{abstract}

\section{Introduction}

    Economists have long studied ownership patterns~\cite{aminadav2020corporate,aziani2020our}, tax haven use by multinational corporations~\cite{desai2006demand,gumpert2016multinational,alstadsaeter2018owns}, profit-shifting \citep{huizinga2008international,dharmapala2019profit}, or how ownership concentration and cross-ownership affects welfare and risk-taking behaviour~\cite{de1999corporate,davis2008new,he2017product,galeotti2021cross}. 
    All this requires foreknowledge of the ultimate controlling entities in the global ownership network. Even though equity ownership is an observed variable (cash-flow rights over firms are recorded in public registers), entity control (the ultimate ability to influence the firm's decision-making) has to be inferred from the latent power relations in the corporate ownership network~\citep{crama2013power}.
    
    A striking universal characteristic of the corporate ownership network is a~bow-tie structure or an~onion-like topology~\citep{glattfelder2009backbone,vitali2011network}. It features the dense core of cross-owned entities surrounded by consecutive layers of nodes with a~lower degree. Such a structure and communities therein reflect the concentration of ownership and power~\citep{vitali2014community,vitali2011geography}. For network science it exemplifies a real-world system with an explicit hierarchical core-shell organisation. 

    In this paper we introduce $\alpha$-ICON (Indirect Control in Onion-like Networks) --- a tuneable algorithm for the identification of the direct/indirect controlling entities that builds on the onion-like (core-shell) properties of the ownership network~\cite{hebert2016multi}. We approach the problem of control in the paradigm of the network flow, also known as ``control by transitivity'': the weights on incident edges are simply multiplied by the given connectivity of the oriented ownership graph. Continuous interpolation between direct and indirect influence is achieved via a damping factor (penalty) $\alpha$~for the length of the flow-paths, inspired by Katz centrality~\cite{katz}. 
    
    
    We exploit the properties of the onion-like topology to propose a time-efficient back-propagation algorithm providing dramatic acceleration for the task of ultimate control identification. To simplify such a topology, we peel off the hanging vertices in the network recursively until a subgraph of nodes with at least one in- and one out-edge remains. This procedure is equivalent to the 2-core decomposition for undirected graphs~\cite{batagelj2002generalized,dorogovtsev2006k}. This is a robust approach that demonstrates the existence of the bow-tie motif~\cite{vitali2011network}. The core-shell decomposition results in a small core in real-world ownership networks, four orders smaller than the whole network. Small core is the fundamental feature of the control network due to restrictions on circular ownership in some jurisdictions. 
    
    Our decomposition naturally sorts all the entities into four fundamental classes, reflecting their topological role in the network: super-holders (nodes that are not owned by any other nodes), super-targets (nodes that do not own any other nodes), intermediaries (nodes that own and are owned by other nodes), and core entities. Specifically, super-holders and super-targets are removed at the first step of our iterative peeling procedure, the intermediaries get removed during the following steps, while the core nodes represent entities that are persistent to the peeling. We note that the ultimate owners (i.e. beneficial owners,~\cite{openownership2020}) are topologically equivalent to super-holders in the control network, thus, the problem of the ultimate control identification is addressed on the class of super-holders.
    
    
    We apply $\alpha$-ICON to the universe of 8.1~mln UK~entities and their controlling participants from the People with Significant Control (PSC) register maintained by the UK Government. This network has only 517 entities at its core, enabling us to identify the ultimate controlling entities much faster than the randomised approximations previously suggested in the literature.
    
    Finally, we evaluate the performance of $\alpha$-ICON against the existing approaches to identify the controlling entities. Public companies have to disclose the information on their controlling interest in subsidiary organisations to the regulatory agencies. We gather such disclosures of the UK companies with debt or securities traded on the US financial markets to build an authoritative list of 1007 subsidiaries and their 279 ultimate controlling entities. This list allows us to compute the recall~@~$k$ of any algorithm: how many subsidiary organisations have the actual controlling entity among the top-$k$ controlling entities identified by the algorithm. We show that $\alpha$-ICON is superior to the previous approaches in terms of recall at any $k$. This advantage is due to efficient detection of super-holder parents located far away from their super-target subsidiaries in the network. To the best of our knowledge, this is the first formal evaluation of the controlling entity identification algorithms. We invite further research by open-sourcing our data and code.\footnote{\url{http://github.com/eusporg/alphaicon}}

\section{Methods}

    \subsection{Unifying direct and indirect control}

        Two literatures have emerged to identify the controlling entities in a~directed graph of participant-entity relations where edge weights are the corresponding equity shares. Statistical physicists consider the flow of power and control in such graphs by studying the topological importance of nodes~\citep{bardoscia2021physics}: centrality~\cite{bovet}, PageRank~\cite{masuda,brin}, coreness~\citep{kitsak2010identification,garas2012kshell}, integrated ownership~\citep{brioschi1989risk}, network control~\citep{vitali2011network}, control capacity~\citep{jia2013control}, control contribution~\citep{zhang2019control}, Helmholtz-Hodge potential~\citep{kichikawa2019community}. Another literature coming from operations research and political science measures effective control in ownership networks by treating them as coalitional games and computing power indices of entities~\citep{shapley1954method,crama2007control}. Randomised approximations such as the Network Power Index (NPI) have recently appeared and are applied to large-scale ownership networks~\cite{mizuno2020power}. The existing approaches can be sorted into one of the two classes: they either rely on the direct ownership or aim at the computation of indirect control propagating \textit{via} all possible routes in the graph of the direct ownership. Even though the indirect measures are considered to better capture the  real distribution of control in the ownership networks, it is still unclear how to balance the contribution coming from the direct and indirect ownership.

        \subsubsection{Direct control}
        
            The equity (ownership) network can be represented as an oriented weighted network. In this model organisations and their participants resemble $N$ nodes while the weights of $E$ edges between them reflect the corresponding equity shares. The edges are oriented due to the non-symmetric nature of the organisation-participant relationship: each edge $\left(i,k\right)$ is oriented from participant $i$ to organisation $k$ and has the corresponding weight $0\le \mathbf{W}_{ik}\le 1$. Normalisation of the shares requires that $\sum_i \mathbf{W}_{ik} = 1$ for any organisation $k$. By construction, we do not allow direct self-ownership, $\mathbf{W}_{ii}=0$. Matrix $\mathbf{W}$ is a non-symmetric adjacency matrix, corresponding to the oriented weighted network.
        
            Control over an organisation can be defined as voting or cash flow rights that a participant has in an organisation. As the matrix $\mathbf{W}_{ij}$ reflects the distribution of direct ownership among the players, it is typically used to approximate the direct control $\mathbf{A}_{ij}$. At the foundation of this model lies the one-share-one-vote assumption~\cite{goergen}. Even though this assumption is broadly accepted in the literature, several other approaches based on transformations on the original matrix $\mathbf{W}$ to the quorum threshold and the relative control were proposed in~\cite{vitali2011network}.
        
            It is worth mentioning a conceptually separate group of models of the direct control coming from the game theory~\cite{crama2007control}. These approaches rely on the assessment of the influence potential among the shareholders and are based on calculation of the power indices in coalitions, such as Shapley-Shubik or Banzhaf \cite{shapley1954method,prigge}. These indices were originally proposed to assess the relative importance of the amount of votes held by each player in a weighted voting system. However, these methods have been criticised in the corporate control literature, since they do not account for widespread heterogeneity of shareholders distribution~\cite{aminadav2020corporate} and due to issues in handling cycles in shareholdings~\cite{crama2013power}. Even though no game-theoretic model has reached universal adoption in the corporate finance literature, there is no denying that such models provide a self-consistent and straightforward algorithm to take into account consolidation of voting rights among direct shareholders.

            In what follows we juxtapose the two approaches in our quantitative model of the direct control: (i) the original matrix of normalised equity shares $\mathbf{A}=\mathbf{W}$ and (ii) the matrix of the Direct Power Indices (DPI) being the randomised approximation of the Shapley-Shubik weights, $\mathbf{A}=\mathbf{W}^{DPI}$. 
    
        \subsubsection{Indirect control}
    
            It is well-known that direct relationships are not sufficient to reveal hidden participants with ultimate control over organisations. Thus, one should take into account the indirect influence a participant~$i$ might exert on the organisation~$j$ through all other participants in the network. In the paradigm of control by transitivity, the indirect control is proportional to the net flow from $i$ to $j$ given the network connectivity. This can be formally defined as the sum of the powers of the adjacency matrix $\mathbf{A}$ of the direct control
                \begin{equation}
                    \sum_{l=1}^{\infty} \mathbf{A}^l = \mathbf{A} + \sum_{k} \mathbf{A}_{ik}*\mathbf{A}_{kj} + \sum_{k,l} \mathbf{A}_{ik}*\mathbf{A}_{kl}*\mathbf{A}_{lj} + \ldots,
                    \label{net_cont}
                \end{equation}
            where the first term is the contribution from the direct control, the second term is the contribution of paths of length 2, the third term is the contribution of paths of length 3 etc. The metric in Eq.~\ref{net_cont} was proposed in~\cite{vitali2011network} for calculation of the network control. However, there are three main issues with this approach.
        
            First, the series in Eq.~\ref{net_cont} is divergent for matrices with the spectral radius $\rho\left(\mathbf{A}\right) \ge 1$. In particular, this is the case for the stochastic matrices of the direct control, $\sum_i \mathbf{A}_{ij}=1$ for each $j$, which have the spectral radius $\rho\left(\mathbf{A}\right)=1$ according to the Gershgorin circle theorem~\cite{gersh}. Thus, in strongly connected components of the ownership network one has to heuristically solve the issues with cycles, a problem which has become classical in the field of the corporate control~\cite{crama2007control,vitali2011network}.
            
            Second, in cases when Eq.~\ref{net_cont} is non-singular, the calculation of control involves inverting the matrix $\mathbf{I}-\mathbf{A}$, which is a computationally demanding task for the whole network of several millions of entities. The authors in~\cite{vitali2011network} suggested a complicated auxiliary algorithm to separate the network into several strongly connected components, which is found to be not efficient when applied to the bow-tie topologies of real-world ownership networks.
        
            Third, the metric of indirect control by Eq.~\ref{net_cont} intrinsically assumes that all paths allowed by the network connectivity contribute equally to the final control, including the paths of length $l=1$ corresponding to the direct control. Still, there is a debate in the field about a quantitative measure that would be optimally balanced between direct and indirect contributions of control~\cite{babic}. In this regard, Eq.~\ref{net_cont} is an absolutely indirect measure being an opposite to the pure matrix $\mathbf{A}$ of the direct control (the first term only).
        
            Here we propose a modification of Eq.~\ref{net_cont} that aims at resolving the three aforementioned issues. We introduce a measure that builds on the ideas of the Katz centrality~\cite{katz} and is defined as follows
            \begin{equation}
                \mathbf{T} = \sum_{l=1}^{\infty} \alpha^{l-1} \mathbf{A}^l = \mathbf{A} \left(\mathbf{I}-\alpha \mathbf{A}\right)^{-1}, \quad \alpha<\rho\left(\mathbf{A}\right)^{-1}.
                \label{katz}
            \end{equation}
            The Katz-like network control in Eq.~\ref{katz} weighs the paths of length $l$ with the factor $\alpha^{l-1}$, where $0\le \alpha<1$ is a tuning parameter of the model, penalising long paths. Thus, this measure continuously interpolates between direct and indirect control. For $\alpha=0$ the control is purely direct, $\mathbf{T}=\mathbf{A}$, while in the limit $\alpha \to 1$ the weights of all possible paths in the network are equal and the purely indirect control Eq.~\ref{net_cont} is recovered. Note that for stochastic matrices $\mathbf{A}$ the series Eq.~\ref{katz} is convergent for any $\alpha<1$. Therefore, we (i)~circumvent the problem of cycles by taking into account the infinite number of paths with exponentially small weight assigned to the long paths and (ii)~propose a metric that continuously interpolates between the direct and indirect control, allowing to fine-tune the model for the particular system under study.
    
        \subsubsection{Core-shell decomposition}
        
            Matrix inversion in~Eq.~\ref{katz} is not computationally efficient in large-scale ownership networks with millions of nodes. We suggest a recursive approach that is based on the core-periphery decomposition of the network. 
        
            Note that the control $T_{ij}$ of $i$ over $j$ can be propagated to $k$, a shareholder of $i$, by increasing the length of each path from $i$ to $j$ by one. The propagated value would be the portion of control of $k$ over $j$ (through the node $i$). If one rewrites Eq.~\ref{katz} as
            \begin{equation}
                \mathbf{T} = \sum_{l=1}^{\infty} {\tilde{\mathbf{T}}}^{(l)}; \quad {\tilde{\mathbf{T}}}^{(l)}=\alpha {\tilde{\mathbf{T}}}^{(l-1)} \mathbf{A},
                \label{katz1}
            \end{equation}
            then Eq.~\ref{katz1} becomes a recursive expression for the control ${\tilde{\mathbf{T}}}^{(l)}$ over paths of certain length $l$. Using Eq.~\ref{katz1}, one can propagate the control upstream from organisations to participants.
        
            Effective propagation of control is conditioned on the absence of cycles in the graph. Following this idea, we propose a procedure similar to the $2$-core decomposition for non-oriented graphs \cite{batagelj2002generalized,dorogovtsev2006k}. Specifically, we iteratively remove the nodes and all the incident edges if the node has only in-going or only out-going edges, i.e. either in-degree $k_{in}^{i}=0$ or out-degree $k_{out}^{i}=0$. At the very first step we remove the participants that are not owned by anyone ($k_{in}=0$) and organisations that do not own anyone ($k_{out}=0$). We classify such entities as super-holders (SH) and super-targets (ST), respectively. Such removal of nodes at the first step results in formation of new ``hanging'' vertices with $k_{in}*k_{out}=0$, which have previously been linked to super-holders and super-targets. After several iterations we converge to a subgraph $C$ of nodes that have at least one in-going and at least one out-going edge. By analogy with the $k$-core definition, the subgraph $C$ can be called $1,1$-core, since this is the maximal subgraph in the network with at least one in-going and one out-going edge at every node. We call this subgraph $C$ ``the core'' which, by definition, is filled with the core entities. Finally, all other nodes in the graph which were removed at step 2 onward are called the intermediaries (I).
            
            \paragraph{Example}
            Consider a weakly connected subgraph in Figure~\ref{fig:network_example} as an example. At the first iteration all green (super-holders) and blue (super-targets) nodes are peeled off from the network. After this iteration the two intermediaries, ``Document Outsourcing Group'' and ``OPUS Property Finance Limited'', become the effective super-targets and are to be removed at the second step. However, the network still possesses hanging vertices: the intermediaries ``OPUS Trust Marketing Limited'' and ``OPUS 107 Limited'' are peeled off at the third and at the fourth iteration steps, respectively. After four iterations the network is left with four organisations for which $k_{in}*k_{out} \ne 0$ (marked orange). They form the $1,1$-core of the network. 
        
            \paragraph{Complexity}
            The described decomposition is of $\mathcal{O}(N)$ computational complexity and ensures absence of cycles in the subgraph of non-core entities (the shell: super-holder, super-targets and intermediaries), allowing for back-propagation of control along all possible paths from the core to the shell. Notably, the size of the core is four orders of magnitude smaller than the size of the whole network for the UK~ownership network we consider in this paper. This allows for dramatic acceleration of the algorithm: while the matrix $\mathbf{T}$ can be straightforwardly computed for the core (demanding only $\mathcal{O}(|C|^3)$ in the worst-case scenario of a dense core), the effect of the rest of the network is calculated via back-propagation to the shell. In this procedure we iteratively compute the control of the intermediaries over the core organisations unfolding them layer by layer in the reverse direction using transitive relations in Eq.~\ref{katz1}. Control over the intermediaries and super-targets, in turn, is computed simultaneously as they appear in the matrix $\mathbf{T}$.
        
            The computational complexity of the back-propagation step is of order of the number of edges in the shell, which is well-approximated by the total number of edges in the ownership network under study, $\mathcal{O}(E)$. Real-world ownership networks reside in the sparse regime (there is of order of one edge attached to each node on average), so the complexity is $\mathcal{O}(N)$. The total complexity of the control calculation is $\mathcal{O}(N+|C|^3)$, where both terms are of the same order for the network considered in the paper.

            \begin{figure}[ht]
                \begin{center}
                	\caption{Principal steps in the $\alpha$-ICON algorithm}
                    \label{fig:diagram}
                	\includegraphics[width=\textwidth]{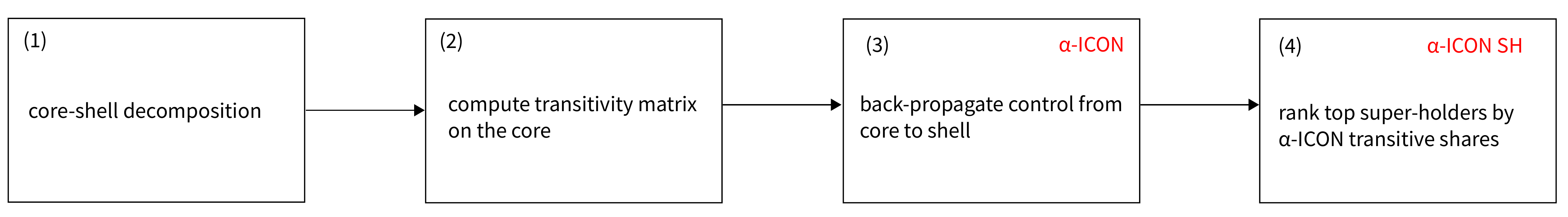}
                \end{center}
            \end{figure}
            
        \subsubsection{Identifying ultimate controlling entities}
        
            Having estimated $\mathbf{T}$ given $\alpha$ for all organisations in the network using back-propagation from the core, we can rank every direct and indirect participant in every organisation by its $\alpha$-ICON share stored in the matrix $\mathbf{T}$. The ultimate controlling entity, by the nature of the ultimate control, is the super-holder with the largest $\alpha$-ICON share. Importantly, these shares are computed as a sum over all possible paths through intermediaries, thus, super-holders cannot be analysed independently without considering their control channelling through other entities in the network.
            
            In order to identify the ultimate controlling entities for each organisation we sort its super-holders by their $\alpha$-ICON shares. If no super-holder is present among the direct or indirect participants we consider any other entity with the largest share. The outcome of this step --- a list of length~$k$ with top-$k$ controlling entities --- is the final result of our algorithm which we designate $\alpha$-ICON~SH (see Figure~\ref{fig:diagram}). 
    
    \subsection{Data}
        
        \paragraph{PSC}
        The UK government maintains a public register identifying the people who can influence the firms' decision-making. Every entity, legal person, or individual holding over 25\% of shares or voting rights in the company or having the power to appoint or remove the majority of the board of directors has to be reported to Companies House, an executive agency of the government~\citep{pscrequirements}. This register, called People with Significant Control (PSC), was established by the Small Business, Enterprise, and Employment Act 2015. It is the companies' responsibility to report and update the information on its direct owners: they are required to file a confirmation statement at least once a year~\citep{heidemann2019commercial}.
        
        Companies House makes the PSC register publicly available. We use its August 2, 2021 snapshot and merge it with the basic company data for live companies (also coming from Companies House) to consider only active (non-dormant) companies in our analysis. PSC register data has gaps in unique participants identifiers. When a participant in organisation is a UK entity we restore its company number by searching for its name in the snapshot of active firms, accounting for namesakes. After data cleaning and removing 282 instances of self-ownership (when the company and its participant are the same entity) we are left with information on ownership of 4~199~626 active companies by 4~042~087 direct participants (including 4~843~147 individuals, 407~224 entities, and 5~971 legal persons) forming $E=5\,256\,342$ participant-entity links in the ownership network. Our $\mathbf{W}\in\mathbb{R}^{N\times N}$ with $N=8\,087\,056$.
        
        An important limitation of the PSC data is that equity shares are reported as intervals: $\left[25\%\ldots50\%\right)$, $\left[50\%\ldots75\%\right)$, $\left[75\%\ldots100\%\right]$. We use midpoints of those intervals and normalise them to $[0...1]$ range for each organisation. Even though it might be worthwhile to treat this share as an interval censored random variable instead, we consider the midpoint normalisation due to its simplicity.\footnote{Our normalisation is not innocuous when there are errors in source equity shares data. When a company has a sole owner with 25\% and many minority shareholders it has one entry in the PSC with $\left[25\%\ldots50\%\right)$ interval and its mid point of 37.5\%. This is the theoretical minimum, and after normalisation this owner will have a share of 1. The theoretical maximum total equity sum per company reaches 150\% when it has 4~owners with 25\% each, or 4~instances of $\left[25\%\ldots50\%\right)$ interval reduced to its midpoint 37.5\%. However, we have 25~160 companies in PSC data with total equity share exceeding 150\%, with the maximum of 2562.5\% (SIMON MIDCO LIMITED). We ignore those errors in summation and normalise equity shares to one weighing by their midpoints.}

        \paragraph{SEC filings}
        Even though several algorithms to identify ultimate controlling entities exist, no formal evaluation has been performed to date, to the best of our knowledge. Primary reason are data requirements. Any evaluation implies knowing the identities of the true ultimate owners of organisations. Gathering such information is labour-intensive. In this paper we benefit from the fact that organisations with debt or securities traded on the US financial markets have to disclose material information to the US Securities and Exchange Commission. In particular, the companies are required to list all their subsidiaries, directly or indirectly controlled, in Exhibit 21 of their SEC 10-K filings. We use those lists from most recent filings (for 2020 and 2021) gathered by CorpWatch~\citep{harwell2007corpwatch}, to identify 1362 parent--subsidiary relations between 304 parents located anywhere in the world and their 1328 subsidiaries --- active UK companies present in Companies House data.
        
        In theory, all such parent-subsidiary relations should exist in the PSC network since all UK companies are required to disclose their ultimate owners. However, we fail to establish 355 parent-subsidiary paths from the SEC data in the PSC network, meaning that the parent and the subsidiary are not connected in the network even though the subsidiary itself exists.\footnote{There are two reasons for such discrepancy. First, we matched 10-K Exhibit 21 filings with PSC data on entity names (excluding namesakes). This procedure may fail to identify the ultimate owners, especially outside the UK, when major differences in the names provided to the SEC and to the PSC arise. Second, companies may intentionally over- or under-report the information to the SEC or Companies House.} After ignoring the unreachable paths we end up with the authoritative list of 1007 subsidiaries and their 279 ultimate controlling entities. Importantly, the distribution of the parent-subsidiary path lengths $l$ on the list is non-degenerate. Only 391 of 1007 subsidiaries (39\%) are directly held by their parents ($l=1$), and 209 subsidiaries (21\%) are more than 3 intermediaries away from their parent ($l > 3$). The longest relation in our evaluation data is $l=12$, meaning that the subsidiary is 11 intermediaries away from its parent.
        
        We use the constructed list to compute the recall~@~$k$ of each algorithm: how many subsidiary organisations have the true parent among the top-$k$ controlling entities identified by this algorithm. The recall ranges from 0 (meaning that the algorithm did not return the correct parent for any of the 1007 subsidiaries among its top-$k$ holders) to 1 (when all the subsidiaries have the correct parent among the top-$k$ holders suggested by the algorithm). The recall can be also decomposed by path length $l$: how many of the subsidiaries $l-1$ intermediaries away from its parent have its true parent among the top-$k$ ultimate controlling entities suggested by the algorithm. This decomposition is useful in comparing the performance of algorithms for the short (direct, $l=1$) and long (indirect, $l>1$) ownership relations.

\section{Results}

    \subsection{Decomposition of the UK ownership network}
    
        We start with the decomposition of the UK's People with Significant Control ownership network into the core and the shell. Namely, we aim at identification of $1,1$-core, which is the maximal subgraph with all vertices having at least one in-going edge and one out-going edge. Clearly, such a subgraph contains cycles, since one can walk along the edges of such subgraph infintely long, while the subgraph is finite. In contrast, the rest of the network is a subgraph without cycles due to the recursive peeling procedure (see Methods). 
    
        The described iterative procedure turns out to be swiftly convergent and results in isolation of the core in 12 iterations. At the very first iteration the largest component with 98\% of nodes disassembles: $N_{SH}=3\,887\,370$ nodes with no in-going edges (super-holders) and $N_{ST}=4\,044\,969$ nodes without out-going edges (super-targets) in the network. In the remaining 11 steps $N_{I}=154\,140$ nodes are removed, despite them having $k_{in}\ge 1$ and $k_{out}\ge 1$ in the original network (intermediaries), see the summary statistics in Table~\ref{tab:summary_stat_by_type}. 
    
        It turns out that the 1,1-core is populated with just 517 companies forming a set of isolated connected components (see Figure~\ref{fig:core_plot}). The absolute majority (84\%) of the core entities are dimers, i.e. companies directly shareholding each other. There are 219 dimers at the core of the British ownership network; we call such components ``trivial'' and do not report them in the Figure~\ref{fig:core_plot}. The smallest and most widely represented (9.3\%) non-trivial component is a triad, in which three companies are connected by at least 3 oriented edges. We also observe 5 tetrads (4.6\%), four of which form a simple 4-cycle and the last one is a hybrid of one triad and one dimer, connected together at one node (Unipart Group of Companies). The largest components are a pentad (Ice Futures) and a hexad (Alpha Financial Markets Consulting), accounting for 2.1\% of the core in sum. All these isolated components for the UK network are strongly connected, i.e. each company controls any other company within the same component. However, in principle, core components might be be weakly connected in some other networks. The smallest theoretically possible weakly connected component of the 1,1-core is a tetrad in which two dimers are connected by a directed link. However, we do not observe such structures in the PSC ownership network.
    
        Sectoral distribution of companies turns out to be quite different at the core compared to the intermediaries or super-targets. 30.6\% of companies at the core are in the Professional, scientific and technical activity sector, while in the universe of the UK companies only 13.6\% are in this sector, a~2-fold difference. Almost 24\% of intermediaries belong to the Financial And Insurance Activities sector, while only 3.4\% of all the UK companies operate in this sector, see Table~\ref{tab:summary_stat_by_type}. Finance/insurance sector together with Professional/Science activities comprise almost 50\% of the core and have a similar share among the intermediaries. At the same time, among the super-targets the respective two sectors comprise only 17\% of organisations (which is close to the share among all live companies in the country).
    
        The lion's share of super-holders are individuals (98.2\%). Since it is impossible for an individual to be a super-target/intermediary or at the core, this is not surprising. An inherent feature of an ownership network is presence of many direct super-holder-super-target relations where super holder is an individual and super-target is an entity. When we compare the mean age or sectoral distribution of super-targets with that in the universe of UK companies in Table~\ref{tab:summary_stat_by_type} we see that they are very close, confirming the ubiquity of a~simple super-holder-super-target relation.
        
    
        \begin{table*}
            \caption{Summary statistics of the UK's People with Significant Control register}
            \label{tab:summary_stat_by_type}
            \begin{threeparttable}
            \resizebox{\textwidth}{!}{\begin{minipage}{\textwidth}
            \begin{tabular}{lSSSSSS}
\toprule 
 & \multicolumn{4}{c}{Position in ownership network} & {\multirow{2}{*}{All companies}} & {\multirow{2}{*}{Evaluation set}}\tabularnewline
\cmidrule(lr){2-5} 
 & {Core} & {Intermediary} & {Super-holder} & {Super-target} & & \tabularnewline
\cmidrule(lr){2-2} \cmidrule(lr){3-3} \cmidrule(lr){4-4} \cmidrule(lr){5-5} \cmidrule(lr){6-6} \cmidrule(lr){7-7}
Entity age, years & 14.58 & 13.32 & 16.97 & 7.55 & 8.41 & 20.46 \tabularnewline
Is public limited company, \% & 0.58 & 0.64 & 0.03 & 0.06 & 0.12 & 2.15 \tabularnewline
Average in-degree & 1.25 & 1.36 & 0 & 1.25 & 1.25 & 0.80 \tabularnewline
Average out-degree & 1.84 & 1.99 & 1.27 & 0 & 0.08 & 0.99 \tabularnewline
\# Entities/individuals, incl.: & 517 & 154140 & 3887370 & 4044969 & 4970026 & 1286\tabularnewline
\indentintable Individual, \% & 0 & 0 & 98.23 & 0 & 0 & 0\tabularnewline
\indentintable Profess./Science activ, \% & 30.56 & 19.24 & 0.07 & 14.19 & 13.63 & 20.14 \tabularnewline
\indentintable Trade, \% & 4.06 & 5.18 & 0.02 & 13.11 & 12.12 & 4.43 \tabularnewline
\indentintable Construction, \% & 4.45 & 7.47 & 0.02 & 10.97 & 10.23 & 0.78 \tabularnewline
\indentintable IT, \% & 4.45 & 5.23 & 0.02 & 9.11 & 8.57 & 8.94 \tabularnewline
\indentintable Admin./Support activ, \% & 14.31 & 8.67 & 0.04 & 8.78 & 8.48 & 8.63 \tabularnewline
\indentintable Real estate, \% & 2.90 & 9.97 & 0.03 & 8.22 & 7.98 & 1.40 \tabularnewline
\indentintable Not available, \% & 10.44 & 4.42 & 1.33 & 4.08 & 6.39 & 25.58 \tabularnewline
\indentintable Other service, \% & 3.09 & 2.22 & 0.03 & 4.81 & 4.75 & 1.32 \tabularnewline
\indentintable Hospitality/Food, \% & 0.77 & 1.90 & 0.01 & 4.86 & 4.70 & 0.31 \tabularnewline
\indentintable Manufacturing, \% & 5.03 & 4.43 & 0.02 & 4.78 & 4.58 & 8.86 \tabularnewline
\indentintable Health/Social work, \% & 0.58 & 2.09 & 0.02 & 3.86 & 3.70 & 0.31 \tabularnewline
\indentintable Finance/Insurance, \% & 17.02 & 23.95 & 0.09 & 2.76 & 3.37 & 17.19 \tabularnewline
\indentintable Transportation/Storage, \% & 0.39 & 1.03 & 0 & 3.45 & 3.30 & 0.86 \tabularnewline
\indentintable Arts/Entertainment, \% & 0.39 & 1.26 & 0.02 & 2.28 & 2.35 & 0 \tabularnewline
\indentintable Households as employers, \% & 0 & 0.17 & 0 & 1.11 & 2.20 & 0 \tabularnewline
\indentintable Education, \% & 0.77 & 0.71 & 0.02 & 1.72 & 1.79 & 0 \tabularnewline
\indentintable Agriculture, \% & 0 & 0.69 & 0 & 0.83 & 0.78 & 0 \tabularnewline
\indentintable Water/Sewerage/Waste, \% & 0 & 0.33 & 0 & 0.34 & 0.34 & 0 \tabularnewline
\indentintable Electricity/Gas/Steam, \% & 0 & 0.63 & 0 & 0.32 & 0.31 & 0.08 \tabularnewline
\indentintable Mining/Quarrying, \% & 0.39 & 0.34 & 0 & 0.21 & 0.21 & 1.17 \tabularnewline
\indentintable Public admin/Defence, \% & 0.39 & 0.05 & 0 & 0.18 & 0.17 & 0 \tabularnewline
\indentintable Extraterritorial orgs, \% & 0 & 0.02 & 0 & 0.02 & 0.03 & 0 \tabularnewline
\bottomrule
\end{tabular}
            
            \footnotesize \textit{Note}: this table reports the means or percentages of entity/individual characteristics by node type in the People with Significant Control register as of August 2, 2021. ``All companies'' are all the live companies in the Companies House register as of August 1, 2021. ``Evaluation set'' are the parent or subsidiary companies forming the set of subsidiaries and their respective parents in the evaluation data.
            \end{minipage}}
            \end{threeparttable}
        \end{table*}
        
    \subsection{Identification of ultimate control}
        
        In order to determine the ultimate owners in the PSC network, we make use of the core-shell decomposition and calculate the $\alpha$-ICON shares of indirect transitive control with almost no damping of possible paths, $\alpha=0.999$ (computation of the control for $\alpha=1$ is not possible due to the divergence of the series Eq.~\ref{katz}). As described in the Methods, first we explicitly calculate the control matrix of the core (square, $517 \times 517$) by Eq.~\ref{katz} and then we iteratively add the nodes from consecutive layers of the shell in the reverse order and compute their control over existing companies in the control matrix. 
    
    \begin{table*}
            \caption{Top-25 super-holders with significant control in the United Kingdom}
            \label{tab:top_holders}
            \begin{threeparttable}
            \resizebox{\textwidth}{!}{\begin{minipage}{\textwidth}
            \begin{tabular}{lS[table-format = 2.0]S[table-format = 3.0]S[table-format = 7.0]S[table-format = 4.1]S[table-format = 4.1]S[table-format = 4.1]S[table-format = 4.0]S[table-format = 3.0]S[table-format = 4.0]S[table-format = 3.0]S[table-format = 1.0]}
\toprule 
Super-holder with significant control & \multicolumn{3}{c}{Rank} & \multicolumn{3}{c}{Sum of controlled shares} & \multicolumn{5}{c}{Network size by node type}\tabularnewline \cmidrule(lr){2-4} \cmidrule(lr){5-7} \cmidrule(lr){8-12} 
 & {$\alpha$-ICON} & {NPI} & {DPI} & {$\alpha$-ICON} & {NPI} & {DPI} & {Total} & {SH} & {ST} & {I} & {C} \tabularnewline \cmidrule(lr){2-2} \cmidrule(lr){3-3} \cmidrule(lr){4-4} \cmidrule(lr){5-5} \cmidrule(lr){6-6} \cmidrule(lr){7-7} \cmidrule(lr){8-8} \cmidrule(lr){9-9} \cmidrule(lr){10-10} \cmidrule(lr){11-11} \cmidrule(lr){12-12}
{\small{}SPECSAVERS OPTICAL SUPERSTORES LTD} & 1 & 1 & 42441 & 1482.2 & 1453 & 4 & 1767 & 188 & 1159 & 420 & 0 \tabularnewline
{\small{}MR PETER VALAITIS} & 2 & 2 & 1 & 1392 & 1392 & 1391 & 1393 & 1 & 1390 & 2 & 0 \tabularnewline
{\small{}PARTNERS GROUP HOLDING AG} & 3 & 3 & 2 & 1166.5 & 1166.5 & 1164.5 & 1170 & 2 & 1167 & 1 & 0 \tabularnewline
{\small{}MR AVTAR SINGH} & 4 & 4 & 4 & 1051 & 1051 & 1051 & 1055 & 3 & 1052 & 0 & 0 \tabularnewline
{\small{}MR ANDREW DAVIS} & 5 & 6 & 2788 & 881.3 & 869.5 & 14.5 & 1020 & 107 & 895 & 18 & 0 \tabularnewline
{\small{}MR ROBERT JARRETT} & 6 & 5 & 5 & 873 & 873 & 873 & 874 & 1 & 873 & 0 & 0 \tabularnewline
{\small{}LLOYDS BANKING GROUP PLC} & 7 & 8 & 24872 & 735.6 & 670.5 & 5 & 956 & 35 & 596 & 325 & 0 \tabularnewline
{\small{}MR CHRIS HADJIOANNOU} & 8 & 7 & 6 & 730 & 730 & 730 & 731 & 1 & 730 & 0 & 0 \tabularnewline
{\small{}BRIDGEPOINT GROUP PLC} & 9 & 10 & 5149 & 669.3 & 591.5 & 11 & 808 & 13 & 536 & 259 & 0 \tabularnewline
{\small{}MR READE GRIFFITH} & 10 & 9 & 491331 & 657.5 & 636.5 & 1.5 & 868 & 25 & 374 & 469 & 0 \tabularnewline
{\small{}PERSIMMON PLC} & 11 & 11 & 2911 & 586 & 578.4 & 14 & 617 & 4 & 561 & 52 & 0 \tabularnewline
{\small{}INTERTRUST N.V.} & 12 & 12 & 887094 & 519.7 & 511.8 & 1 & 530 & 4 & 353 & 173 & 0 \tabularnewline
{\small{}IWG PLC} & 13 & 13 & 10 & 481.8 & 481.7 & 422 & 497 & 6 & 473 & 18 & 0 \tabularnewline
{\small{}BARRATT DEVELOPMENTS P L C} & 14 & 17 & 587 & 481.1 & 473.6 & 31 & 530 & 9 & 446 & 75 & 0 \tabularnewline
{\small{}FRASERS GROUP PLC} & 15 & 16 & 2982 & 480.1 & 474.8 & 14 & 484 & 2 & 438 & 44 & 0 \tabularnewline
{\small{}MR GRZEGORZ SZEWCZYK} & 16 & 14 & 8 & 480 & 480 & 480 & 481 & 1 & 480 & 0 & 0 \tabularnewline
{\small{}MR DARREN SYMES} & 17 & 15 & 9 & 475.5 & 475.5 & 474.5 & 478 & 2 & 475 & 1 & 0 \tabularnewline
{\small{}PETS AT HOME GROUP PLC} & 18 & 19 & 1226949 & 468 & 389.1 & 1 & 770 & 216 & 514 & 40 & 0 \tabularnewline
{\small{}BGF GROUP PLC} & 19 & 23 & 6219 & 462.1 & 343 & 10 & 674 & 99 & 378 & 197 & 0 \tabularnewline
{\small{}OCTOPUS CAPITAL LIMITED} & 20 & 18 & 1624 & 414.5 & 411.7 & 19 & 431 & 4 & 310 & 117 & 0 \tabularnewline
{\small{}M\&G PLC} & 21 & 20 & 17408 & 404 & 386.6 & 6 & 469 & 14 & 298 & 157 & 0 \tabularnewline
{\small{}BP P.L.C.} & 22 & 26 & 1757 & 377.8 & 336.4 & 18 & 493 & 5 & 390 & 96 & 2 \tabularnewline
{\small{}IVC ACQUISITION PIKCO LTD} & 23 & 21 & 1535931 & 370 & 366.3 & 1 & 371 & 1 & 341 & 29 & 0 \tabularnewline
{\small{}SUN LIFE FINANCIAL INC.} & 24 & 22 & 2086664 & 364.7 & 358.2 & 1 & 469 & 10 & 242 & 217 & 0 \tabularnewline
{\small{}PLACES FOR PEOPLE GROUP LIMITED} & 25 & 25 & 458 & 345.8 & 338.7 & 35.5 & 387 & 11 & 324 & 52 & 0 \tabularnewline
\bottomrule
\end{tabular}
            
            \footnotesize \textit{Note}: this table reports the 25 super-holders with the largest total $\alpha$-ICON transitive share in all entities when $\alpha=0.999$. Also reported are the sums and ranks of their respective NPI and DPI indices. We also provide the information on the topological types of nodes sharing the same sub-network with this super-holder. Ownership data comes from the People with Significant Control register as of August 2, 2021.

            \end{minipage}}
            \end{threeparttable}
        \end{table*}
    
        We also compute the NPI and DPI indices~\cite{mizuno2020power} for the entire ownership network using our own implementations of those algorithms.\footnote{Quota = 0.5, 10~000 iterations. For NPI we discarded the first 1000 iterations and set original label restoration probability $\epsilon=0.01$.} As the baseline approach we identify the controlling entities using the direct equity shares, i.e. the original matrix of shares $\mathbf{W}$ before normalisation. We compare the three approaches to the results of $\alpha$-ICON algorithm by examining the top-25 super-holders in Table~\ref{tab:top_holders}. 
    
        Surprisingly, the two sets of top-25 holders in the ownership network determined by their $\sum\alpha$-ICON and $\sum$NPI --- another algorithm of indirect control --- closely match each other. Furthermore, the first four super-holders are ranked the same by the two methods. Interestingly, the first four super-holders (but the first one) have equal total sums of shares by $\alpha$-ICON and NPI. The leading super-holder, Specsavers Optical Superstores Ltd, is predicted to control roughly 30 companies more by $\sum\alpha$-ICON than by $\sum$NPI. Even though NPI does identify fewer subsidiary companies held by the Specsavers, this does not change the ranking order. Note, however, that further down the list lower $\sum$NPI in relation to $\sum\alpha$-ICON does change the rank of the subsequent holders. 

        Naturally, the $\sum$DPI ranks outlined in Table~\ref{tab:top_holders} are much less compatible with $\sum\alpha$-ICON (when $\alpha=0.999$) or $\sum$NPI, since DPI provides a measure of direct control. Still it is instructive to cross-compare the results of the two approaches. The second, the third and the fourth super-holders by both $\sum\alpha$-ICON and $\sum$NPI have $\sum$DPI shares with almost identical absolute values of control. This means that these super-holders exert control over their subsidiaries directly, without intermediaries. In contrast, the first super-holder has the $\sum$DPI rank 42~441, meaning that it has a distributed network of subsidiaries not accounted for by DPI. We depict a similar difference between direct and indirect control in Figure~\ref{fig:network_example}(b) where the super-holder The Berkeley Group PLC directly controls 39 entities and 177 entities in total, if we include indirect control.
    
        Another case in point is Mr Reade Griffith, ranked 10th by $\sum\alpha$-ICON, 9th by $\sum$NPI, and 491~331st by $\sum$DPI. He exerts his control predominately indirectly compared to Mr Peter Valaitis, ranked 2nd by $\sum\alpha$-ICON and $\sum$NPI, and 1st by DPI, who is the only super-holder in his network of control. This is consistent with the classification of entities within the connected components controlled by top-ranked super-holders in the Table~\ref{tab:top_holders}. Mr Peter Valaitis represents the central node in his star-like connected component with 1392 leaves, which correspond to the values of $\sum\alpha$-ICON and $\sum$NPI. However, as we see, $\sum$DPI identifies only 1391 subsidiaries, because one of them is at distance~2 from Mr Peter Valaitis and, thus, is hidden from the algorithms of direct control. On the other hand, in the component of the first holder, Specsavers Optical Superstores Ltd, there are more than twice as many intermediaries as super-holders. Despite it being the primary beneficiary holder in the UK, only~4 out of~1400+~of its subsidiaries are identified by DPI. This is an example of a strongly distributed ownership network in which determination of the leading beneficiary holder is non-trivial. In Figure~\ref{fig:network_example}(a) we plot this network where the super-holder Specsavers Optical Superstores Ltd (in green) owns a large stake in intermediary Specsavers UK Holdings Limited (depicted in grey in the centre) that, in turn, co-owns many super-targets (in blue) alongside many other super-holders (also in green).
    
        Interestingly, the networks of all top-25 super-holders, but one (BP P.L.C.), do not contain cycles. This is reflected in the absence of the core nodes therein. As a rule, the control networks of the main UK super-holders represent trees. From the practical standpoint, this drastically simplifies determination of the ultimate control in these networks.
    
    \subsection{Evaluation}
       
       \begin{figure}
            \begin{center}
            	\caption{Recall of controlling entity identification algorithms}
                \label{fig:evaluation_figure}
        
            	\centering
            	\subfloat[Recall @ $k$ across algorithms]{
            	    \scalebox{0.5}{
            		\includegraphics[width=1\linewidth]{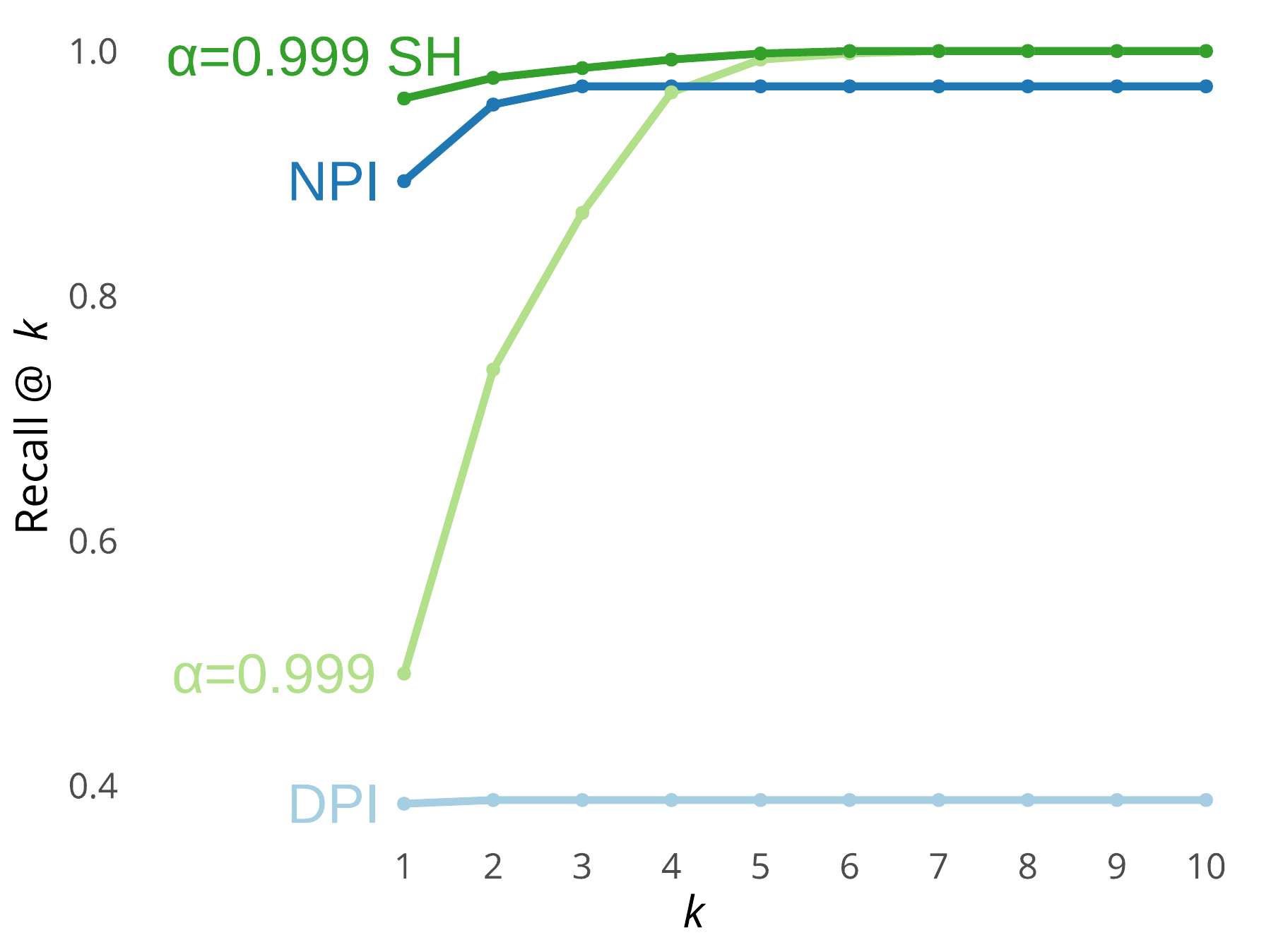}
            		}    	
            	}
            	\subfloat[Recall @ $k$ for $\alpha$-ICON SH, $\alpha \in \left(0.1,\ldots,0.999\right)$]{
            	    \scalebox{0.5}{
            		\includegraphics[width=1\linewidth]{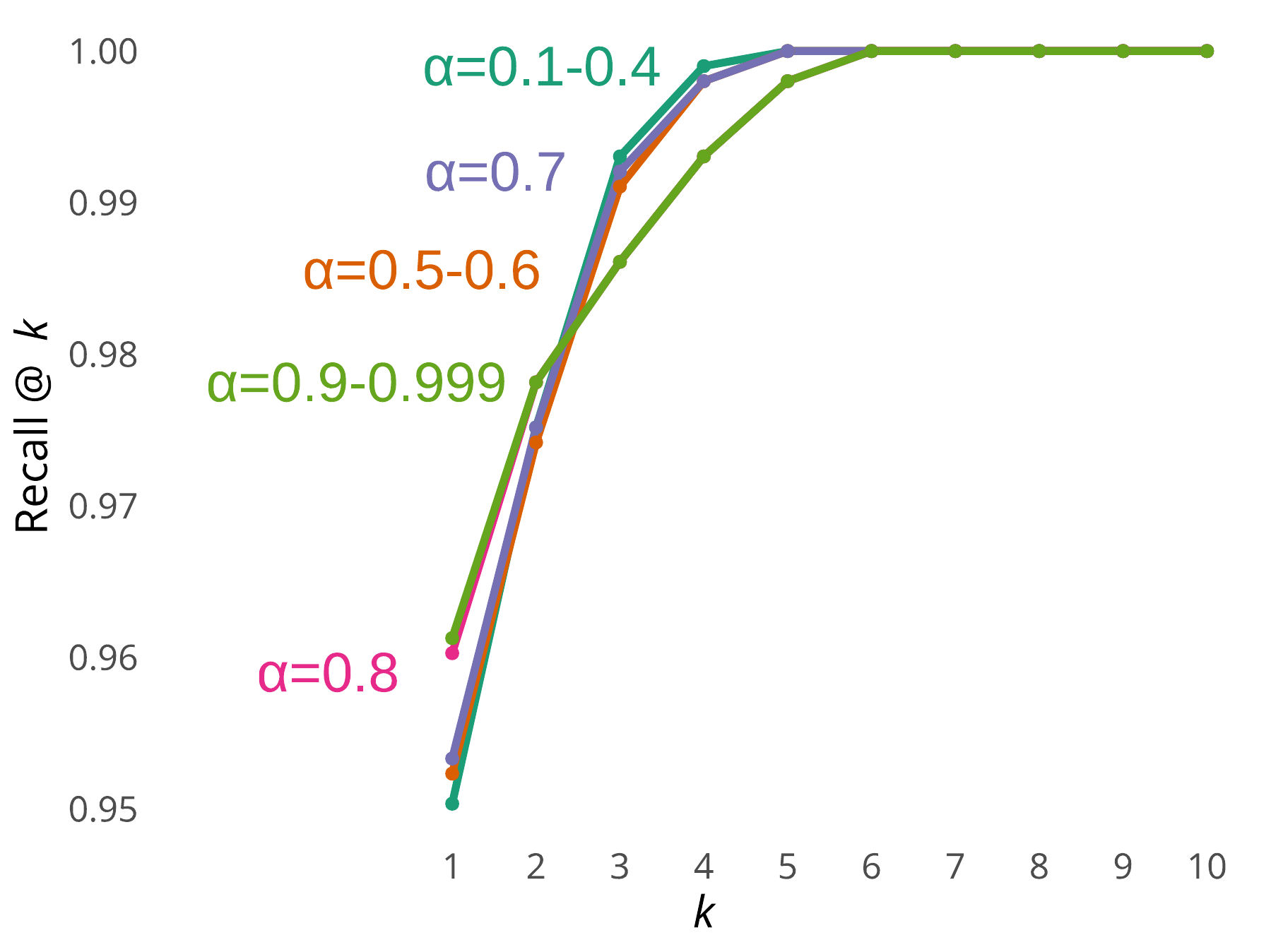}
            		}
            	}
            	
                \subfloat[Recall @ 1 across algorithms by path length]{
            	    \scalebox{0.5}{
            		\includegraphics[width=1\linewidth]{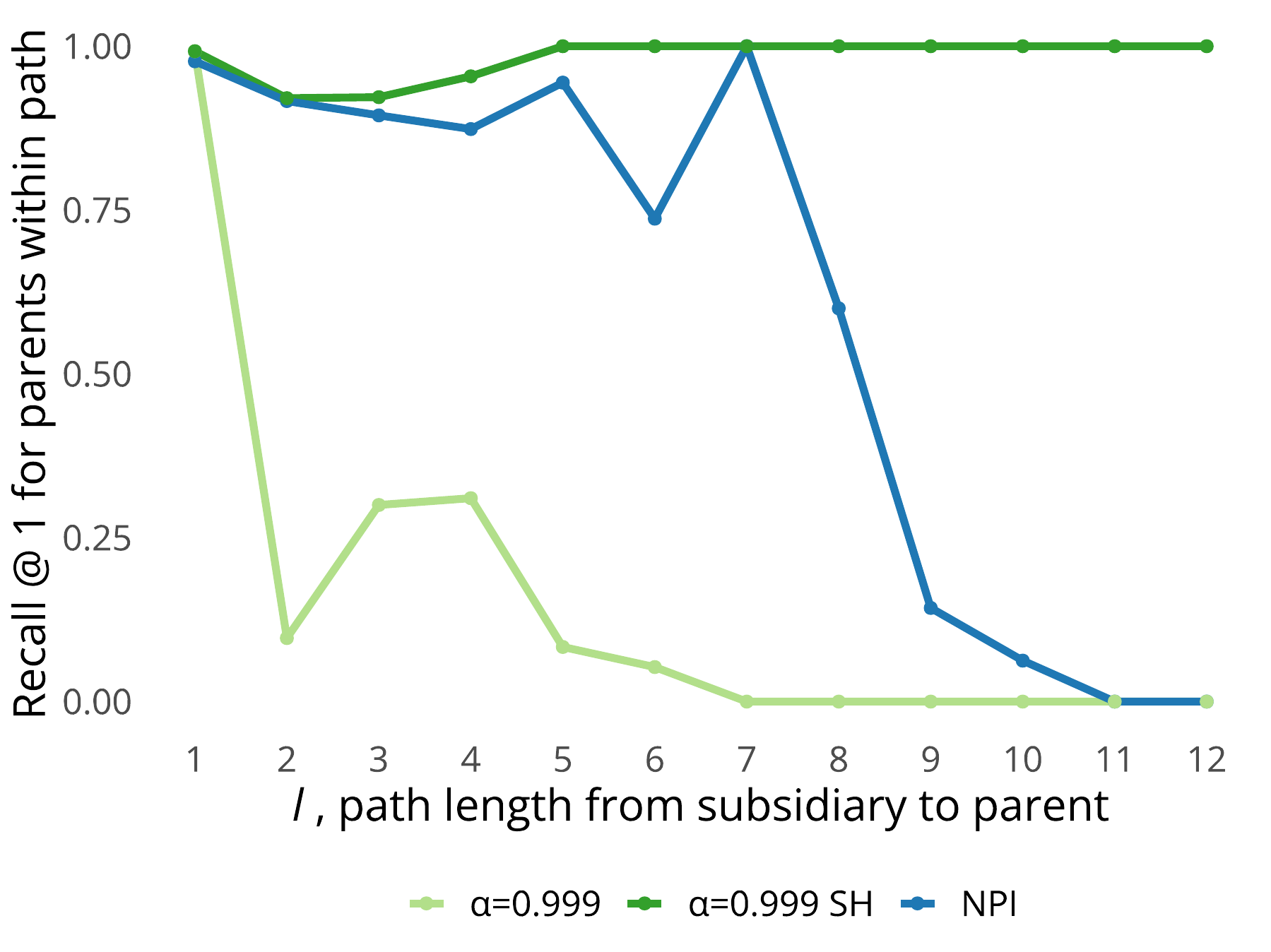}
            		}
            	}
            	\subfloat[Recall @ 5 across algorithms by path length]{
            	    \scalebox{0.5}{
            		\includegraphics[width=1\linewidth]{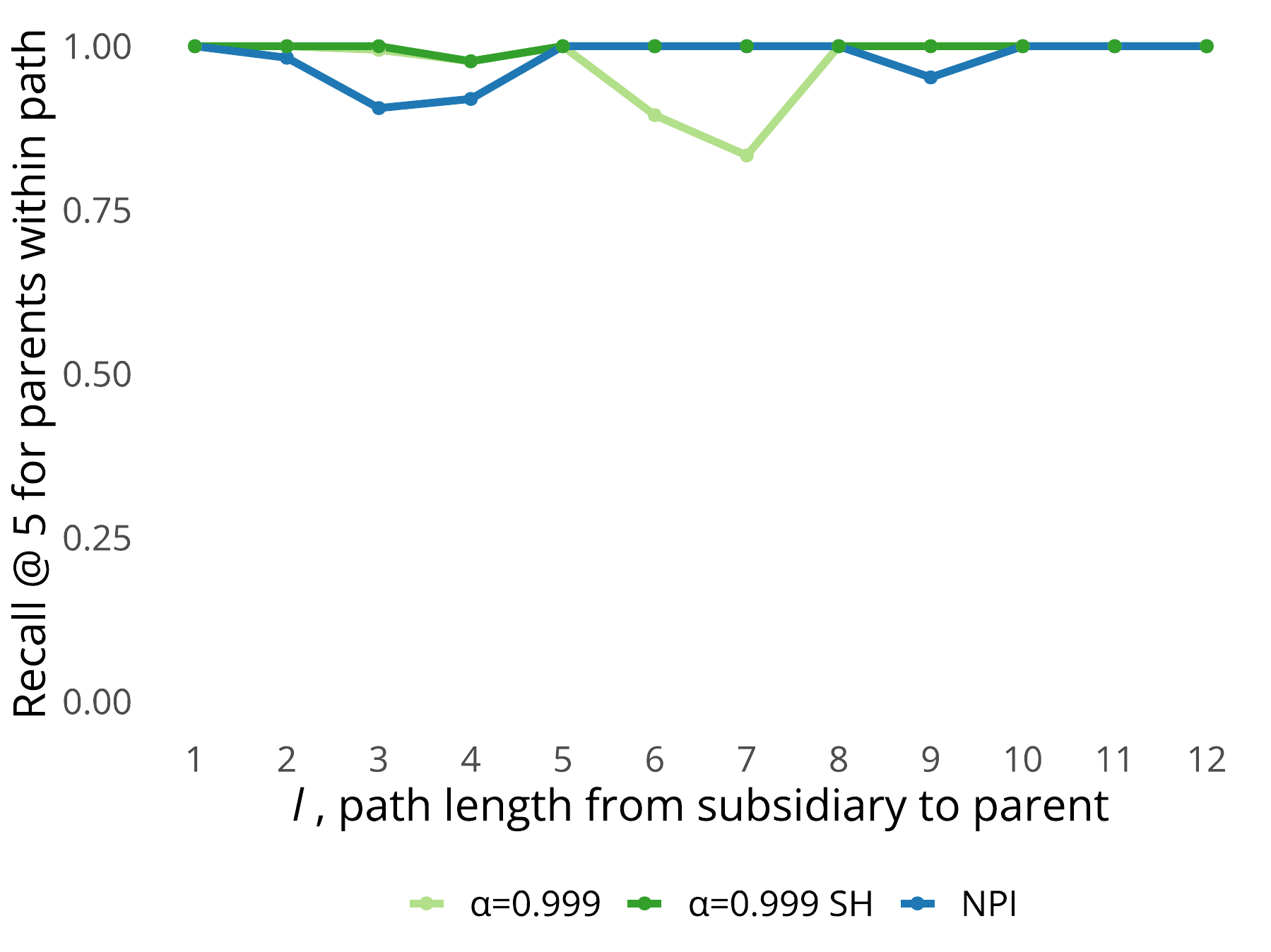}
            		}
            	}
            \end{center}
            \footnotesize \textit{Note}: detailed evaluation results from all models are reported in Table~\ref{tab:evaluation_table}.
        \end{figure}
        
        We engage in an evaluation study to understand the validity of various control identification algorithms using the ground-truth set of parent-subsidiary relations from SEC 10-K Exhibit 21 filings of companies with debt/equity traded on the US financial markets (see Data section). Note that this list is not topologically representative of the whole ownership network or the UK companies (compare column ``Evaluation set'' and ``All companies'' in Table~\ref{tab:summary_stat_by_type}). Notwithstanding, it allows us to compare the control identification algorithms.
    
        We make use of the core-shell decomposition and calculate the $\alpha$-ICON shares of indirect transitive control for a series of $\alpha=0,0.1,0.2,\ldots, 0.8, 0.9, 0.999$. For the quantitative comparison of the algorithms we estimate how many subsidiary companies have the correct parent among its top-$k$ super-holders identified by an algorithm (recall~@~$k$). Results of this computation for different~$k$, $\alpha$ and various algorithms of control are reported in Figure~\ref{fig:evaluation_figure}(a)--(b) and Table~\ref{tab:evaluation_table}(a).
        
        We immediately see from Figure~\ref{fig:evaluation_figure}(a) that DPI (light blue line) with recall of 38.5\% @ 1 performs the worst in the evaluation, highlighting the need to account for indirect ownership. In contrast, $\alpha$-ICON SH (for $\alpha=0.999$, dark green line) is superior to all conventional algorithms and yields 96.1\% of correctly determined ultimate holders at recall~@~1 and 99.8\% at recall~@~5. In particular, the~$\alpha$-ICON SH is superior to NPI (dark blue line): at recall~@~1 the fraction of correctly determined ultimate holders is 6.7 p.p. lower for NPI.
        
        The performance gain of $\alpha$-ICON is due to the final step of our algorithm when we search for entities with the largest share only among the super-holders. Absent this filter, $\alpha$-ICON (light green line) yields worse results. This is expected since the entities with the largest transitive share are the core entities and the intermediaries. As a result, there are owners that exert control over them, despite having lower transitive share in the network. The problem of the ultimate control is different from the identification of the ultimate owner in the way that the former should be analysed in the class of super-holders. When we consider the~$\mathbf{W}^{DPI}$ weights in lieu of the normalised $\mathbf{W}$ equity shares~$\alpha$-ICON SH demonstrates somewhat worse results, Table~\ref{tab:evaluation_table}(a) suggests. However, it still registers better performance than NPI. We conclude that the transitive control computed by~$\alpha$-ICON SH produces more coherent results.
        
        $\alpha$-ICON SH has one tuning parameter. In Figure~\ref{fig:evaluation_figure}(b) we compute its recall at different~$\alpha$s. One could immediately notice non-monotonic relationship between algorithm quality and~$\alpha$ at different $k$. At $k=1$ the highest recall is achieved by $\alpha \in \left[0.9,\ldots,0.999\right]$, with recall for $\alpha=0.8$ being the close second. Lower values of $\alpha$ produce lower recall~@~1. The situation changes starting from recall~@~3. $\alpha<0.8$ produces better results in such settings. This might hint at the existence of optimal damping parameter $\alpha^*$ that maximises the recall.
    
        To further understand the source of discrepancy between the algorithms, it is instructive to look at the evaluation performance at different subsidiary-parent path lengths $l$, see Figure~\ref{fig:evaluation_figure}(c)--(d) and Table~\ref{tab:evaluation_table}(b). $\alpha$-ICON SH algorithm (dark green line) displays monotonically increasing recall~@~1 starting from $l=2$, while the performance of NPI (dark blue line) is clearly not stable with the path length. This analysis suggests that NPI fails to correctly identify ultimate super-holders located at large distance~$l$ from their subsidiaries in the ownership graph. With increase of~$k$ this effect essentially vanishes, Figure~\ref{fig:evaluation_figure}(d) suggests.
    
\section{Discussion}
    
    Our algorithm to determine the ultimate controlling entities in ownership networks scales linearly with network size and supersedes previous approaches in terms of its recall. We observe that $\alpha$-ICON with $\alpha$ very close to unity takes into account all possible paths of control in the ownership network, overestimating the shares of distal super-holders along the ownership graph. At the same time, NPI overestimates the weight of nearby holders by allowing them to form coalitions with other participants. This observation is crucial for the discussion to follow.
    
    \paragraph{Cycles}
    Previous algorithms had to develop \textit{ad hoc} procedures to overcome cycles induced by circular ownership. Our algorithm, in contrast, is built around cycles by isolating them at the network's $1,1$ core. This approach allows for a dramatic speed-up because cycles are rare in real-world ownership networks. In fact, we are surprised to find any cycles in the UK PSC network given that subsidiaries cannot be members of its holding company.\footnote{Companies Act 2006, Section 136(1)(a).} However, previous studies of the PSC data reached similar conclusions regarding the circular ownership \citep{globalwitness2019}. With explicit prohibition of circular ownership in place it is reasonable to expect the UK ownership network to feature a small core, which is indeed the case. The smaller the core, the faster the $\alpha$-ICON algorithm. 
    
    \paragraph{Quota}
    Heuristic approaches and algorithms rooted in game theory require setting multiple parameters, including the quota, or the equity share threshold to influence the firm's decision-making. While most authors opt for the simple majority quota of 50\%, this assumption may dramatically alter the control identification. Consider the example of an ownership network of Boparan Holdings Ltd in Figure~\ref{fig:network_example}(c). This is an involved ownership network super-held by two individuals: Ranjit Boparan (with his 50--75\% stake) and Bajlinder Boparan (with her 25--50\% stake). Since the stake of the latter super-holder is less than the 50\% quota she does not control any firm in the network per DPI or NPI. Conversely, under the transitive ownership of $\alpha$-ICON when $\alpha=0.999$ she has the total control of 24 in comparison to Ranjit Boparan's 73. By lowering $\alpha$ we lower her total control up until it reaches $\nicefrac{37.5}{37.5+62.5}=0.375$ when $\alpha=0$ (direct control only after midpoint normalisation). In principle, one can easily integrate the quota in $\alpha$-ICON by equating edges with weight less than the specified threshold to zero in adjacency matrix $\mathbf{W}$.

    \paragraph{Optimal $\alpha^{*}$}
    The higher the value of $\alpha$ the more weight is given to distant ownership relations. This might be undesirable in networks where there is a legal restriction on the maximal ownership path length. If the maximal path allowed is $l_{max}$, the optimal $\alpha^{*}$ can be assessed from $l_{max}=-\nicefrac{1}{\ln\left(\alpha^{*}+1\right)}$. An alternative strategy is to gather a validation data set with the true subsidiary-parent relationships that represents the network in terms of its topological and non-topological characteristics. Then $\alpha^{*}$ is the value at which the recall reaches maximum in this validation data set. While this is an attractive way forward, it is not deprived of well-known challenges. Gathering a representative validation set is a daunting task when the majority of relationships in real-world ownership networks are simple super-holder-super-target links. In a network populated only with such links we do not need anything more than $\alpha^{*}=0$ (i.e. direct control only) to capture the control in its entirety. In a network where the majority of links are the simple SH-ST but there exists a handful of extremely complicated sub-networks of indirect ownership setting $\alpha^{*}=0$ would result in a severe under-estimation of the influence of indirect players. Therefore, it might be worthwhile to build a validation dataset that accurately represents both direct and indirect ownership in the network. Using this benchmark one could pick $\alpha^{*}$ that results in the highest cumulative average of recalls @ $k$ for every path length $l$. Under this approach one can achieve the balance between penalising long paths and fostering the indirect control.

    \paragraph{Limitations}
    For a dense core network, the computational complexity of $\alpha$-ICON is cubic in the size of the core. In networks with large and dense strongly connected components amounting to tens of thousands of nodes in the core computation of the transitivity might become unfeasible. However, an approximation by several first components of the transitivity matrix (in the spirit of the eigenvector centrality) can significantly reduce the computational complexity. We leave this to future research. Another limitation relates to the interpretation of our findings on control shares. The top-25 super-holders in Table~\ref{tab:top_holders} include the individual holders of thousands of firms, retail chains with complicated structures, and multinational corporations. It is hard to believe that the supermajor oil company BP PLC (market capitalisation of USD~84.1 bln) ranked 22th per $\sum\alpha$-ICON exerts less influence than Pets at Home PLC (market capitalisation of USD~3.5 bln) ranked 18th. Our transitive ownership shares are all but the starting point in determination of the total control. They, for instance, can be used as weights in summation of net assets or market capitalisation of companies held by the super-holders.
    
\section{Conclusion}

    In this paper we have proposed $\alpha$-ICON --- a~robust numerical algorithm for identification of ultimate control in onion-like ownership networks that may include consisting of millions of nodes.

    Our algorithm borrows from the ideas of Katz centrality to calculate the network flow with the additional damping factor, exponentially weighting the probability of long paths of transitive control. In contrast to previous methods, our approach allows for the continuous interpolation between fully direct control (when only direct shareholders can exert their influence) for $\alpha=0$ and purely transitive network control in the limit $\alpha \to 1$ (when control paths of all lengths in the network are equally likely to exist). In order to satisfy the domain constraint ruling out intermediaries from the ultimate holders, our algorithm determines top-$k$ beneficiaries for each company by ranking its super-holders by the computed $\alpha$-ICON shares.

    We are first to offer formal evaluation of controlling entity identification algorithms. We gather such disclosures of the UK companies with debt or securities traded on the US financial  markets to build an authoritative list of subsidiaries and their ultimate controlling entities. We show that $\alpha$-ICON is superior to the previous approaches. In particular, we find that the recently proposed indirect control algorithm, NPI, though being close in performance to $\alpha$-ICON, fails to correctly identify \emph{distal} parents of subsidiaries. This is because NPI is based on particular game-theoretic model of control by consolidation of votes, which is quite a debatable approach in corporate control literature due to heterogeneity in distribution of shareholders~\citep{aminadav2020corporate} and circular shareholding~\citep{crama2013power}.
    
    We reveal the universal core-shell topological organisation of the ownership network with a small core (4 orders of magnitude smaller than the whole network), which allows for dramatic acceleration of calculation of the transitive control. We also note that the widely known bow-tie structure of real-world ownership networks is a reflection of their fundamental degeneracy and sparsity. This inherent property corresponds to the $1,1$-core decomposition ($k_{in}=1$ and $k_{out}=1$) of the oriented graphs. We find that only less than 0.01\% of the whole ownership graph belongs to (in general, weakly, but strongly in practice) connected components with cycles (the core), while the remaining 99.99\% of entities grow on trees (the shell). Our algorithm exploits this feature to supply dramatic optimisation via recursive back-propagation of control from the core to the shell along the tree-like paths. 

    Importantly, the restriction of long transitive paths at $\alpha<1$ suggests a universal approach to the problem of votes consolidation in determination of the ultimate control. While conventional methods taking into account consolidation rely on~a particular game-theoretic model (such as Shapley-Shubik), resulting in the model-specific typical path length between organisation and its beneficiary in the ownership graph, our approach is advantageous as it is capable to explore the beneficiaries at all path lengths (resolutions) simultaneously. We note that the possibility of coalition formation is the particular reason for the transitive network control (at $\alpha=1$) to be not optimal. 

    In general, the optimal value of $\alpha^*$, reflecting some sort of balance between direct and indirect control, should be chosen guided by the data. In this paper, based on a novel evaluation procedure, we demonstrate that the values $\alpha>0.8$ overestimate the role of long paths in the control ranking, though still showing better performance than NPI and DPI. To make any credible conclusions on the optimal $\alpha^{*}$ inherent to the whole network, the evaluation dataset should be topologically and economically representative to the network. In practice, the latter condition is not always possible to satisfy, which, thus, poses an interesting problem for future research.

\section{Acknowledgements}

    Authors are grateful to Interdisciplinary Scientific Center Poncelet, European University at St.~Petersburg, and Skoltech for organisational support at the initial stages of this work. KP and NP acknowledge partial support of RSF grant \texttt{21-11-00215}.

\section*{References}

    \bibliography{bibliography}
    
\clearpage
\appendix


\setcounter{table}{0}
\setcounter{figure}{0}
\renewcommand{\thetable}{S.\arabic{table}}
\renewcommand{\thefigure}{S.\arabic{figure}}
\pagenumbering{Roman}
\renewcommand{\thepage}{App.~\arabic{page}}

\onecolumn

\section*{Supplementary materials} \label{note:Appendix} 

            
                
            
    
    \begin{table}[ht]
    	\caption{Results of evaluation on SEC 10-K Exhibit 21 subsidiaries data}
        \label{tab:evaluation_table}

    	\begin{center}
    	\subfloat[Recall @ top-$k$ ultimate controlling entities]{
    	    \scalebox{1}{\begin{tabular}{lS[table-format = 1.3]S[table-format = 1.3]S[table-format = 1.3]S[table-format = 1.3]}
\toprule 
 & \multicolumn{4}{c}{Recall @ $k$} \tabularnewline \cmidrule(lr){2-5}
 & {1} & {3} & {5} & {10} \tabularnewline 
\cmidrule(lr){2-2} \cmidrule(lr){3-3} \cmidrule(lr){4-4} \cmidrule(lr){5-5}
Baseline & 0.384 & 0.388 & 0.388 & 0.388\tabularnewline
DPI & 0.385 & 0.388 & 0.388 & 0.388\tabularnewline
NPI & 0.894 & 0.971 & 0.971 & 0.971\tabularnewline
NPI, SH & 0.894 & 0.971 & 0.971 & 0.971\tabularnewline
$\alpha$-ICON, $\mathbf{W}$, $\alpha=0$ & 0.384 & 0.388 & 0.388 & 0.388\tabularnewline
$\alpha$-ICON, $\mathbf{W}$, $\alpha=0.999$ & 0.492 & 0.868 & 0.993 & \textbf{1.000}\tabularnewline
$\alpha$-ICON, $\mathbf{W}$, $\alpha=0.999$, SH & \textbf{0.961} & \textbf{0.986} & \textbf{0.998} & \textbf{1.000}\tabularnewline
$\alpha$-ICON, $\mathbf{W}^{DPI}$, $\alpha=0.999$, SH & 0.911 & 0.985 & 0.997 & 0.999\tabularnewline
\bottomrule
\end{tabular}}
        }
        
        \subfloat[Recall @ 1 for subsidiaries $l$ intermediaries away from their parents]{
    	    \scalebox{1}{\begin{tabular}{lS[table-format = 1.3]S[table-format = 1.3]S[table-format = 1.3]S[table-format = 1.3]S[table-format = 1.3]S[table-format = 1.3]S[table-format = 1.3]}
\toprule 
 & \multicolumn{7}{c}{Recall @ 1 for subsidiary-parent path length $l$}\tabularnewline \cmidrule(lr){2-8}
 & {1} & {2} & {3} & {4} & {5} & {10} & {12}\tabularnewline
\cmidrule(lr){2-2} \cmidrule(lr){3-3} \cmidrule(lr){4-4} \cmidrule(lr){5-5} \cmidrule(lr){6-6} \cmidrule(lr){7-7} \cmidrule(lr){8-8} 
Baseline & 0.990 & 0.000 & 0.000 & 0.000 & 0.000 & 0.000 & 0.000\tabularnewline
DPI & \textbf{0.992} & 0.000 & 0.000 & 0.000 & 0.000 & 0.000 & 0.000\tabularnewline
NPI & 0.977 & 0.916 & 0.894 & 0.874 & 0.944 & 0.062 & 0.000\tabularnewline
NPI, SH & 0.977 & 0.916 & 0.894 & 0.874 & 0.944 & 0.062 & 0.000\tabularnewline
$\alpha$-ICON, $\mathbf{W}$, $\alpha=0$ & 0.990 & 0.000 & 0.000 & 0.000 & 0.000 & 0.000 & 0.000\tabularnewline
$\alpha$-ICON, $\mathbf{W}$, $\alpha=0.999$ & \textbf{0.992} & 0.097 & 0.300 & 0.310 & 0.083 & 0.000 & 0.000\tabularnewline
$\alpha$-ICON, $\mathbf{W}$, $\alpha=0.999$, SH & \textbf{0.992} & 0.921 & \textbf{0.922} & \textbf{0.954} & \textbf{1.000} & \textbf{1.000} & \textbf{1.000}\tabularnewline
$\alpha$-ICON, $\mathbf{W}^{DPI}$, $\alpha=0.999$, SH & \textbf{0.992} & \textbf{0.930} & 0.917 & 0.920 & 0.944 & 0.062 & 0.000\tabularnewline \cmidrule(lr){1-8}
\# subsidiaries at path length & {391} & {227} & {180} & {87} & {36} & {16} & {1} \tabularnewline 
\bottomrule
\end{tabular}}
        }
        \end{center}
    \end{table}
    
    \begin{figure}[ht]
        \caption{Core-periphery decomposition of an ownership network}
       \label{fig:network_example} \includegraphics[width=0.99\linewidth]{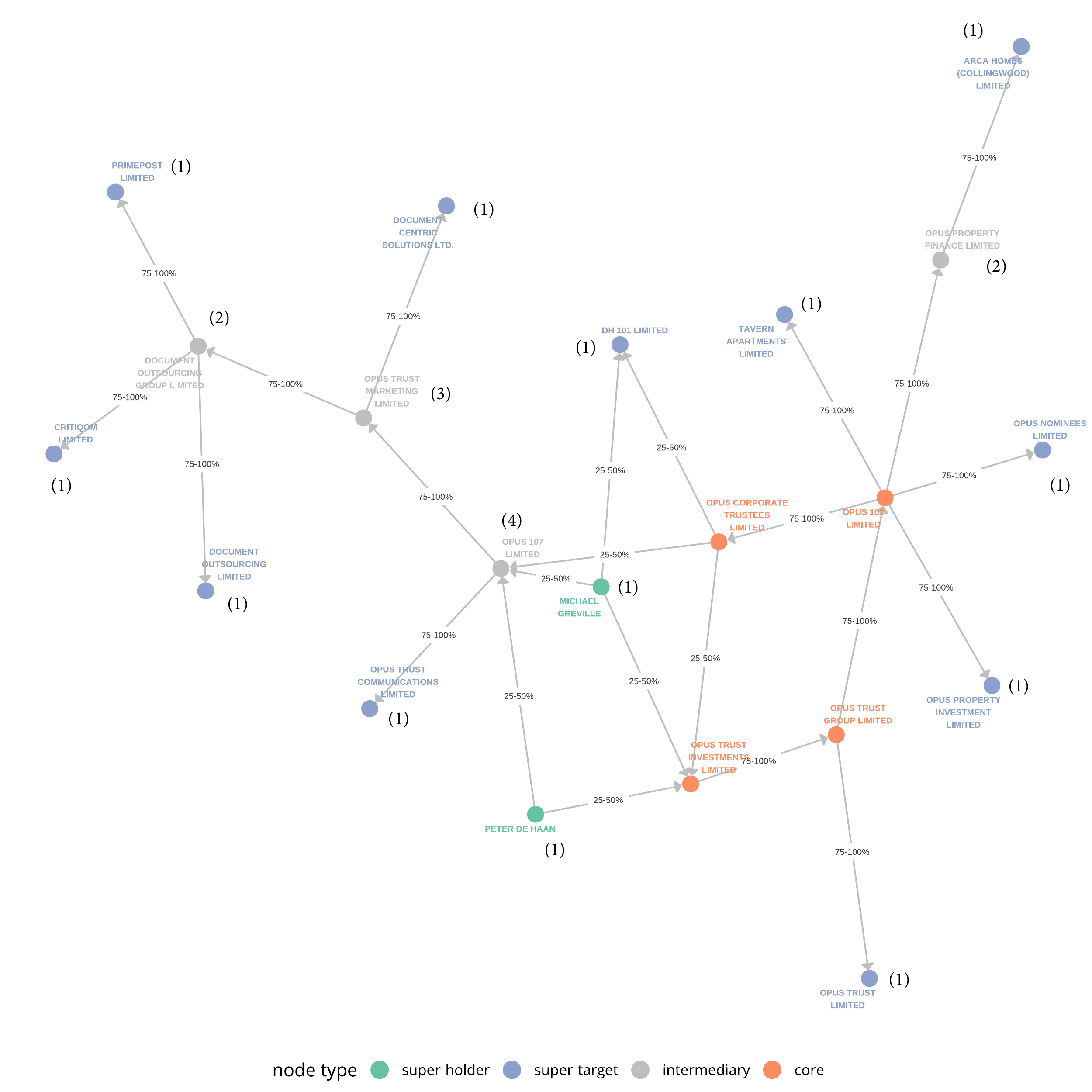}
       
       \footnotesize \textit{Note}: colour codes the inferred node types. Edge texts are equity shares. Numbers in parentheses are iterations when the respective nodes are peeled off. 
    \end{figure}
    
    \clearpage
    
    \begin{figure}[ht]
    \begin{center}
    	\caption{All non-trivial components (excl. dimers) at the core of the UK's People with Significant Control register (node size is proportional to total holdings)}
    	\label{fig:core_plot}

    	\includegraphics[width=\textwidth]{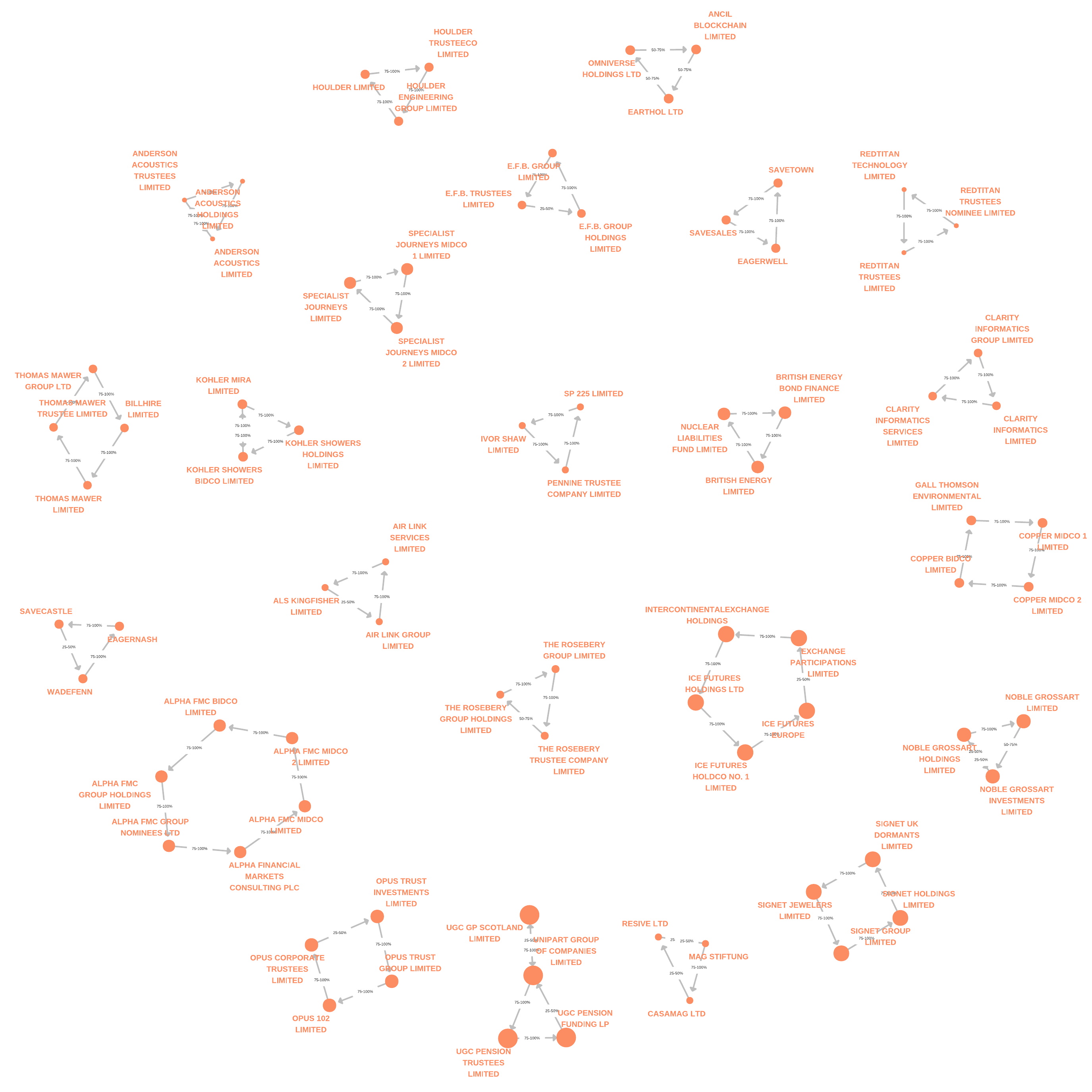}
    \end{center}
    \end{figure}
    
    \clearpage

    \begin{figure}[ht]
    	\caption{Selected ownership networks from the UK's People with Significant Control register (node size is proportional to log of direct holdings)}
    	\label{tab:graph_examples}

    	\begin{center}
    	\subfloat[A network with the largest $\alpha$-ICON SH and the largest difference between direct and indirect control by SH: Specsavers Optical Superstores LTD \footnotesize{($\sum$DPI=4, $\sum$NPI=1453, $\sum\alpha$-ICON=1482)}]{
			\scalebox{1}{
            	\includegraphics[width=0.9\linewidth]{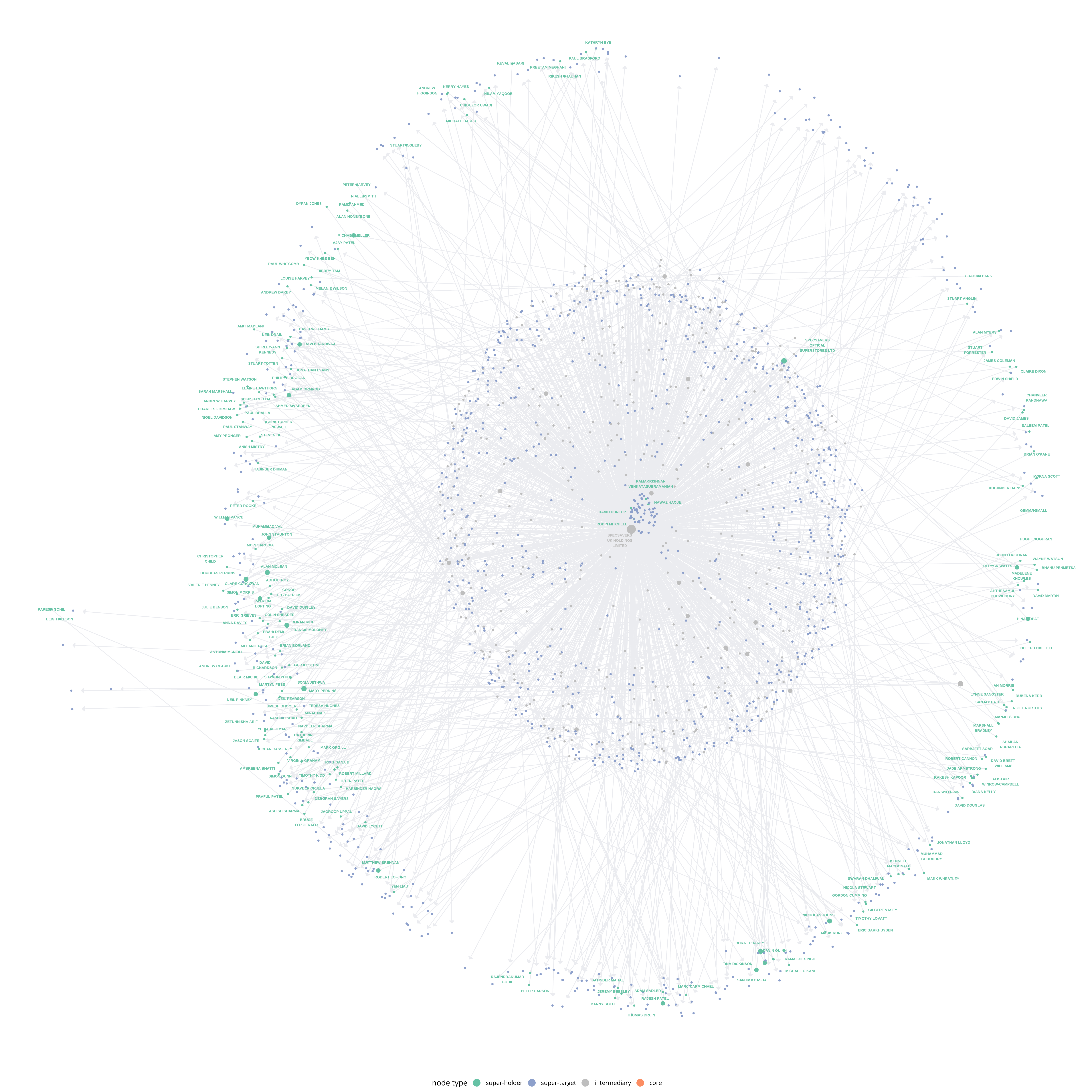}
			}
    	} \\
    	\subfloat[A network where $\alpha$-ICON and DPI differ for~SH:\\The~Berkeley~Group PLC \footnotesize{($\sum$DPI=39, $\sum\alpha$-ICON=177)}]{
		    \scalebox{0.5}{
        		\includegraphics[width=1\linewidth]{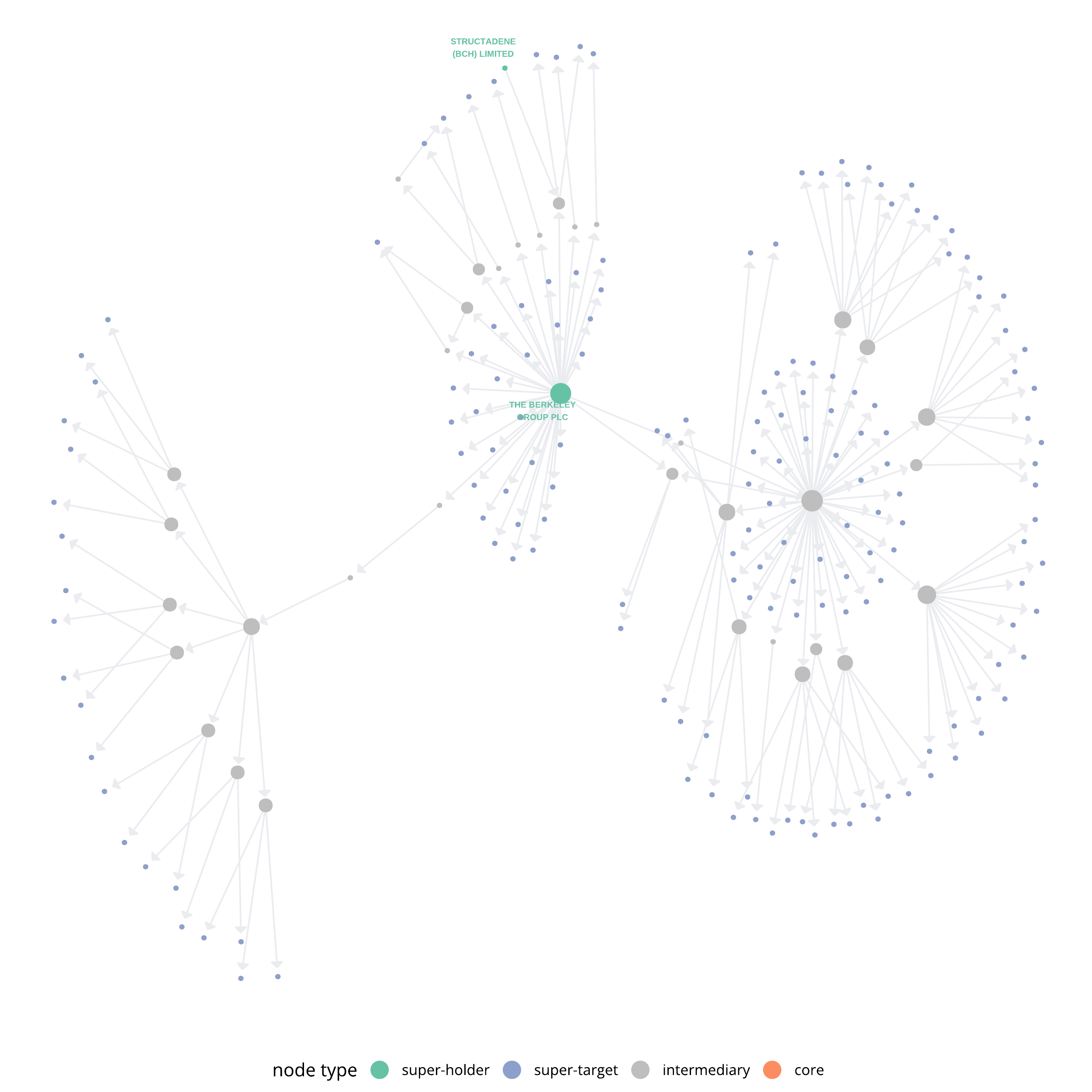}
		    }
    	}
        \subfloat[A network where $\alpha$-ICON and NPI differ for~SH:\\Bajlinder~Boparan \footnotesize{($\sum$NPI=0, $\sum\alpha$-ICON=24)}]{
		    \scalebox{0.5}{
        		\includegraphics[width=1\linewidth]{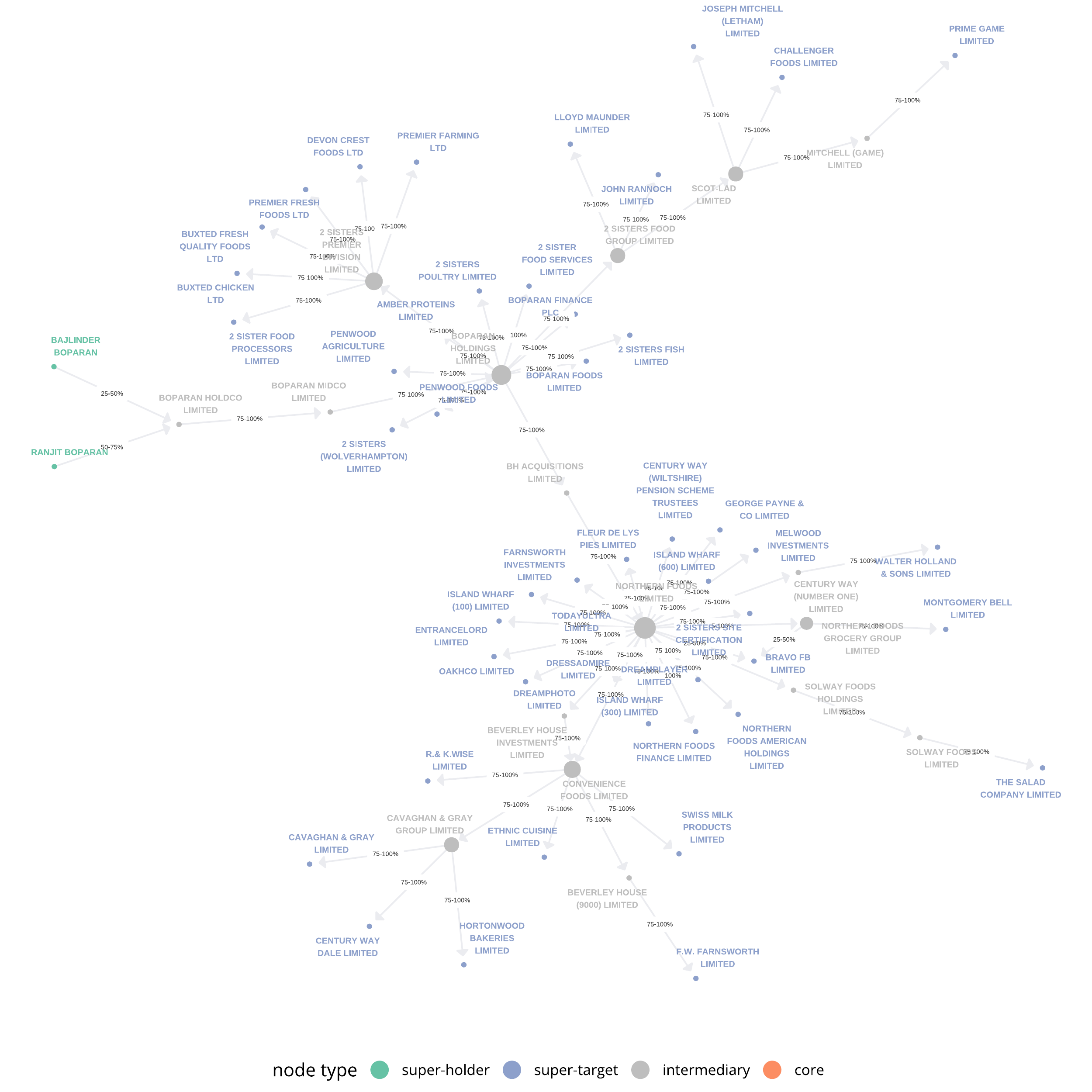}
		    }
    	}\\[10ex]
    	\end{center}
    \end{figure}
    \clearpage
    
    

    %



\end{document}